\documentclass[aps,prd,twocolumn,showpacs,floatfix,preprintnumbers,amsmath,amssymb,nofootinbib,groupedaddress, superscriptaddress]{revtex4-1}
\usepackage[latin1]{inputenc}
\usepackage{amsfonts}
\usepackage{subfigure}
\usepackage{hyperref}
\usepackage{bookmark}
\usepackage{slashbox}
\usepackage{natbib}
\usepackage{dcolumn}
\usepackage{bm}
\usepackage{latexsym}
\usepackage{epsfig}
\usepackage{graphicx,multirow}
\usepackage{units}
\usepackage{color,soul}
\usepackage{amssymb}
\usepackage{amsmath}
\usepackage{enumerate}
\usepackage{textcomp}
\usepackage{mathtools}
\graphicspath{ {figures/} }
\def\re#1{(\ref{#1})}

\newcommand{\be}{\begin{equation}}
\newcommand{\ee}{\end{equation}}
\newcommand{\ba}{\begin{eqnarray}}
\newcommand{\ea}{\end{eqnarray}}

\newcommand{\gsim}{\;\mbox{\raisebox{-0.5ex}{$\stackrel{>}{\scriptstyle{\sim}}$}}\;}
\newcommand{\lsim}{\;\mbox{\raisebox{-0.5ex}{$\stackrel{<}{\scriptstyle{\sim}}$}}\;}

\begin{document}
\title{CMB Spectral Distortions from Cooling Macroscopic Dark Matter}

\author{Saurabh Kumar}
\email{saurabh.kumar@case.edu}
\affiliation{Department of Physics/CERCA/Institute for the Science of Origins, Case Western Reserve University, Cleveland, OH 44106-7079 -- USA}
 
\author{Emanuela Dimastrogiovanni}
\email{exd191@case.edu}
\affiliation{Department of Physics/CERCA/Institute for the Science of Origins, Case Western Reserve University, Cleveland, OH 44106-7079 -- USA}
\affiliation{Perimeter Institute for Theoretical Physics, 31 Caroline St. N., Waterloo, ON, N2L 2Y5, Canada}

\author{Glenn D. Starkman}
\email{glenn.starkman@case.edu}
\affiliation{Department of Physics/CERCA/Institute for the Science of Origins, Case Western Reserve University, Cleveland, OH 44106-7079 -- USA}

\author{Craig Copi}
\email{cjc5@case.edu}
\affiliation{Department of Physics/CERCA/Institute for the Science of Origins, Case Western Reserve University, Cleveland, OH 44106-7079 -- USA}

\author{Bryan Lynn}
\email{bryan.lynn@cern.ch}
\affiliation{Department of Physics/CERCA/Institute for the Science of Origins, Case Western Reserve University, Cleveland, OH 44106-7079 -- USA}
\affiliation{Department of Physics and Astronomy, University College London, Gower Street, London WC1E 6BT, UK}

\date{\today}

\begin{abstract}
We propose a new mechanism by which dark matter (DM) can affect the early and late universe. The hot interior of a \textsl{macroscopic} DM, or \textsl{macro}, can behave as a heat reservoir so that energetic photons and neutrinos are emitted from its surface and interior respectively. In this paper we focus on the spectral distortions (SDs) of the cosmic microwave background before recombination. The SDs depend on the density and the cooling processes of the interior, and the surface composition of the Macros. We use neutron stars as a model for nuclear-density Macros and find that the spectral distortions are mass-independent for fixed density. In our work, we find that, for Macros of this type that constitute 100$\%$ of the dark matter, the $\mu$ and $y$ distortions can be near or above detection threshold for typical proposed next-generation experiments such as PIXIE.
\end{abstract}
\maketitle
\section{Introduction}

In the standard $\Lambda$-CDM model of cosmology, dark matter (DM) comprises $\Omega_{\mathrm{DM}}\sim0.27$ of the total energy density of today's universe. 
From observations of the cosmic microwave background (CMB)
and cosmic structures, 
we know this DM must be ``cold'' ({\it i.e.} non-relativistic),
and ``dark'' ({\it i.e.} interact rarely with ordinary matter and radiation) --
hence  Cold Dark Matter (CDM).  It must also interact rarely with itself.

The microscopic nature of CDM is still unknown;
however, the absence of a suitable  Standard Model (SM)
particle has driven the widespread belief that the DM is 
a Beyond the Standard Model (BSM) particle, 
and a concomitant decades-long search for such particles
in purpose-built DM detectors 
and among the by-products of collisions at particle accelerators.  

The ongoing infertility of such particle DM searches, 
whether for Weakly Interacting Massive Particles (WIMPs) or axions,
suggests that other candidates return to serious consideration.
Two such candidates have long histories: 
so-called ``primordial'' black holes (PBHs), 
and  composite baryonic objects of approximate nuclear density and macroscopic size,
which we will refer to as ``Macros'' \cite{2015MNRAS.450.3418J}, although that term properly
includes macroscopic candidates of any density and composition.

This paper is focused on Macros,
and specifically on  observational consequences of the presence of 
nuclear-density Macros in the early universe.
These have the particular virtue that they may be 
purely SM objects built of quarks or baryons.
In this case they must have been formed before the
freezeout of weak interactions at $t\simeq1s$
and the subsequent onset  of big bang nucleosynthesis (BBN), 
if the success of the
standard theory of BBN in predicting light-element abundances 
is to be preserved (although see \cite{Dimopoulos:1987fz}).

Witten first suggested \cite{Witten:1984rs}
that DM could be 
composites of up, down and strange quarks 
assembled during the QCD phase transition.  
Subsequent proposals have included purely SM
objects made of quarks \cite{strangematter} 
or baryons \cite{strangebaryon,chiral} 
of substantial average strangeness.
A variety of BSM variations also exist
({\it e.g.}, \cite{colorsuperconductor}). 
Several authors have focused on the observational consequences
(\cite{nuclearites,impactor,2015MNRAS.450.3418J}).

Macros share with PBHs 
an important distinction from particle DM candidates:
they achieve their low interaction rates by being massive,
and therefore of much lower hypothetical number density.
Non-observation of approximately nuclear-density Macros through the tracks they would have left in ancient mica
\cite{2015MNRAS.450.3418J} demands $m_X\gsim 55$g.
Limits on larger Macro  masses have been  obtained 
from gravitational micro-lensing ($m_X\lsim 2\times10^{20}$g)
\cite{microlensingMACHO, microlensingGriest, microlensingSUBARU}
and femto-lensing 
(excluding $10^{17}\lsim m_X\lsim 2\times10^{20}$g).
These limits as quoted
assume that the DM consists 
of Macros of a single mass -- 
an unlikely situation for a composite object.
Macroscopic bound states of fermions (e.g. quarks or baryons) 
cannot be formed by  gravitational collapse of adiabatic fluctuations
in the radiation dominated era.
They would arise typically from non-adiabatic fluctuations \cite{Witten:1984rs,1990PhLB..240..179N} or topological defects \cite{Liang:2016tqc} (e.g. from phase transitions)
that enhance the fermion abundance relative to that of the radiation. 
Although there are stringent constraints on kaon or pion condensates, 
hyperons, and strange quark matter inside observed 2-$M_{\odot}$ or 
heavier neutron stars~\cite{2012ARNPS..62..485L,2001ApJ...550..426L},  
these states may (or may not) be found in  lighter neutron stars.  
Moreover, these exotic hadronic or quark matter equations of state (EOS) are theoretically allowed; hence, one should not abandon the possibility of their playing a role in the structure of Macros, which are certainly not the endpoints of ordinary stellar evolution. The mass functions of Macros are model-dependent
and therefore difficult to predict~\cite{1990PhLB..240..179N}.  
We do not discuss further the origin of the Macros, 
but concern ourselves with their detection.

Cosmological constraints on Macros, 
whether from the CMB or large scale structure, 
do not yet impinge on generic nuclear-density objects.

The presence of dense assemblages of quarks or baryons from
before BBN through today 
would undoubtedly have  as-yet unexplored 
observational consequences, no matter the specific mechanism of 
their formation or stabilization.  
Novel physics peculiar to such Macros, 
with potential observational consequences include:\newline
\noindent1) slow pre-recombination cooling of the Macro compared to the ambient plasma:\newline
\noindent\phantom{SPA}a. distorting the spectra of the cosmic microwave \newline
\noindent\phantom{SPAa.}and  neutrino backgrounds (CMB and CNB), \newline
\noindent\phantom{SPA}b. heating the post-recombination universe, or \newline
\noindent\phantom{SPA}c. contributing to the cosmic infrared background;\newline
\noindent2) production of nuclei (including heavy nuclei) through:\newline
\noindent\phantom{SPA}a. inefficiency in Macro assembly at formation,\newline
\noindent\phantom{SPA}b. evaporation, sublimation or boiling, especially \\\indent \textcolor{white}{xxx} soon after Macro formation,\newline
\noindent\phantom{SPA}c. Macro -Macro collisions;\newline
\noindent3)\textcolor{white}{,}formation of binary Macros, with potential gravitational-wave and electromagnetic signals;\newline
\noindent4) DM self-interactions, especially in high-dens{}ity environments such as galactic cores;\newline
\noindent5) enhanced thermal and dynamical coupling of dark-matter to baryons and photons.

These primary processes could have important secondary consequences, including implications for early star formation, assembly of supermassive black holes, and 21-cm emissions \cite{Fialkov}.

In this work, we focus our attention on the very first of these: 
the effect on the CMB and CNB
of macroscopic objects that generically cool
by volume emission of neutrinos and surface emission of photons. 
(BSM candidates may have additional cooling mechanisms.)

By considering a specific example of a baryonic Macro -- 
a neutron star (NS) -- as a Macro, we demonstrate that the weak coupling of neutrinos to baryons
and the inefficiency of surface cooling by photons
generically lead the Macros to remain 
significantly hotter than the ambient plasma
through the epoch of recombination. 
Both energy and entropy are therefore injected into the plasma
in the form of photons and neutrinos
well after the time when thermal or statistical
equilibrium can be restored.
The CMB and CNB spectra are thereby distorted. 

In this first work, we characterize the distortion
in terms of the traditional 
$\mu$-type (photon-number excess) and
$y$-type (photon-energy excess) spectral distortions (SD) 
of the CMB,
and by  $\Delta N_{eff}$, 
the increase in the effective number of neutrino species.
However, because the temperature of the Macros can remain
far above that of the ambient plasma, 
and because the cooling is ongoing through and after recombination,
neither $\mu$ nor $y$ will fully capture the shape of the
resulting distortion.  
This will be considered in future work,
as will the angular fluctuations in the distortion,
its correlation with other observables,
and other potential consequences 
of baryonic Macro DM, 
as listed above.

The magnitude of SD caused by  Macros is controlled 
of course by  their abundance, 
but also by their specific internal physics.  
For NS material this includes:
the thickness and insulating properties of the non-degenerate crust;
the equation of state of the neutrino-emitting core, 
in particular the presence/absence of a superconducting phase 
and its detailed properties.

For a solar-mass NS, 
known or anticipated properties result in $\mu$ and
$y$-type distortions of the CMB
that are potentially above the threshold of detection 
by feasible next-generation SD experiments, and  $\Delta N_{eff}$ that are not.
These specific conclusions will change 
for other microphysical models of Macros, 
but may be instructive of what to expect and why.
To our knowledge, this is the first study of 
the radiative cooling of DM and the
CMB spectral distortions it may cause.

The CMB has been measured \cite{Mather:1993ij} 
to have a black body (BB) spectrum 
with an average temperature of $2.7255\pm0.0006$K. 
Some deviation from a BB is predicted
due to energy injection/absorption mechanisms~\cite{barrowcoles,adiabaticbaryons,1969Ap&SS...4..301Z,1970Ap&SS...7....3S,1977NCimR...7..277D,Hu:1992dc,2012MNRAS.419.1294C},
especially the damping of acoustic modes 
after they have entered the horizon, a.k.a. Silk damping~\cite{1970Ap&SS...7....3S,barrowcoles,1991ApJ...371...14D,1994ApJ...430L...5H,silkdamping,2012PhRvL.109b1302P,2012ApJ...758...76C,2015PhRvD..91l3531E,2016JCAP...12..015D,2017MNRAS.466.2390C}.
At very high redshifts, $z> z_\mu\equiv2\times10^{6}$, distortions would be wiped out 
by efficient photon number and energy-changing interactions. 
For $5\times10^{4}\lesssim z \lesssim2\times10^{6}$, 
number-changing mechanisms are inefficient,
and photon injection results in a finite chemical potential 
in the Bose-Einstein distribution of photons, a
so-called $\mu$ distortion. 
At lower redshifts, $z \lesssim 5\times10^{4}$, 
energy redistribution by Compton scattering becomes inefficient,
leading to $y$-type distortions. 
The intermediate era, $z\approx10^4 - 10^5$, is also characterized by \textsl{i-type} distortions~\cite{2012JCAP...09..016K}.

The only macroscopic objects of nuclear density 
known to exist in nature 
are NS formed as endpoints of stellar evolution.  
These appear to have masses below $2.2M_\odot$ 
	\cite{Margalit:2017dij,Alsing:2017bbc},
well below the total mass within the horizon at $z\sim 10^{9}$
(or even $10^{12}$, the epoch of quark confinement and chiral 
symmetry breaking). 
We therefore use an ordinary NS as a proxy for a Macro.
We take the Macro's central density to be 
$\rho_{\mathrm{X}}
\simeq\rho_N\equiv2.8\times10^{14} \mathrm{g/cm^{3}}$,
which we refer to as nuclear density. 
Although microlensing limits preclude 
	all the DM  being NS, the Macro mass function 
	could include a sizable contribution from them.

Neutron stars theoretically are stable down to 
	$(0.09-0.19)M_\odot$ \cite{1993ApJ...414..717C,2006IJMPA..21.1555B,Potekhin:2013qqa,Belvedere:2013hla},
but do not appear to arise 
as the endpoints of the evolution of main-sequence stars
below  $\sim1.2M_\odot$.
If formed in the early universe, 
these would be larger and of lower average density 
than post-stellar neutron stars.
This motivates us to consider NS-like Macros 
of somewhat lower-than-nuclear density.

The discovery of a NS with $M_{NS}<1.2M_\odot$
would be exciting evidence for early-universe Macro formation.

Smaller-still composite baryonic objects 
require non-gravitational stabilization, 
whether within the SM 
through the incorporation of strange quarks/baryons~\cite{Witten:1984rs},\cite{strangebaryon},\cite{chiral} 
or by more exotic BSM mechanisms. 
Such SM or BSM baryonic composites 
may also exist in the mass range that includes stable NS.

The paper is organized as follows. In Section~\re{macro_cooling}, we discuss the neutrino and photon 
emission processes that cool the Macro. In Section~\re{macros_sd}, we calculate the SD created by photon 
emission from the surface of Macros. In Section~\re{conclusions}, we present our conclusions. We provide a 
derivation for the photon luminosity of the Macro, and describe the neutrino cooling processes in more detail in the appendices.

\section{Cooling of Macros}
\label{macro_cooling}

In this section, we provide expressions for the neutrino and photon luminosities from the interior 
and surface of the Macro respectively. We then arrive at the temperature dependence of the interior 
of the Macro. 

Except for the inner core, 
the composition of which is still under debate, 
a NS is composed 
of neutrons, protons, electrons and heavy ions. 
After formation, it cools down via  
neutrino emission from the interior 
and photon emission from the surface. 

Neutrino cooling occurs through three processes:\newline
\noindent1) direct Urca (DUrca) 
	\[
	n \rightarrow p + e^{-} + \bar{\nu}_{e}\,,\,\,\, e^{-} + p \rightarrow  n + {\nu}_{e}
	\]
	takes place at high temperatures, 
	when neutrons and electrons are non-degenerate,
	but may also be important below the degeneracy temperature;
\newline
\noindent2) modified Urca (MUrca) 
in the neutron and proton branches
\begin{eqnarray} 
n + n &\rightarrow n+ p + e^{-} + \bar{\nu}_{e},\,\, 
   n + e^{-} + p \rightarrow  n + n + {\nu}_{e} \nonumber\\
n + p &\rightarrow p+ p + e^{-} + \bar{\nu}_{e},\,\,\, 
   p + e^{-} + p \rightarrow  n + p + {\nu}_{e}\nonumber
\end{eqnarray} 
is dominant at  $T < 10^{9}$K,
when neutrons and electrons are degenerate;\newline
\noindent3) nucleon Cooper pair (CP) cooling
\[ \tilde{N} + \tilde{N} \rightarrow \text{CP} + \nu + \bar{\nu},\]
	(where $\tilde{N}$ is a quasi-nucleon)
	is most efficient for 
	$0.98 T_{c}\gsim T \gsim 0.2 T_{c}$, 
	with neutrons in the NS interior 
	become superconducting at $T_c$~\cite{nsreview}.    

The luminosity of these neutrino-cooling processes is 
\begin{equation}
  \label{luminosity}
\mathcal{L}^i_{\nu} = 
	10^{45} C_i 
	\left(M_{\mathrm{X}}/M_\odot\right)
	\left(\rho_{\mathrm{X}}/\rho_N\right)^{-1/3}
	 \mathrm{erg/s}
\end{equation}
for $i=DU,MU,CP$. $M_{\mathrm{X}}$ is the mass of the Macro. $\rho_{\mathrm{X}}$ is the density of the core, and it partly characterizes our ignorance of the precise properties of the Macro.  
$T$ dependence is encoded in  
\begin{equation}
        C_i =
        \begin{cases}
        5.2 ~(T_{9}^{X})^6~R^{D}&  i=DU \nonumber\\
                (3.0R^{M}_{{n}}+2.4R^{M}_{{p}}) 10^{-6} ~(T_{9}^{X})^8~\alpha&
                                i = MU \nonumber\\
                7.1\times10^{-6}\left(\rho_{X}/\rho_{N}\right)^{-1/3}  ~(T_{9}^{X})^7~a ~F&  i = CP\,. \nonumber
  \end{cases}
\end{equation}
$T^{X}$ is the Macro's internal temperature;
the subscript $9$ will be used 
for a temperature in units of $10^9$K. In the above equation, $\alpha\simeq2\left(1+m_{\pi}^{2}/p_{F}^{2}(n)\right)^{-2}- 0.3\left(1+m_{\pi}^{2}/p_{F}^{2}(n)\right)^{-1} + 0.07$ where $p_{F}(n) = 340 (\rho_{\mathrm{X}}/\rho_{\mathrm{N}})^{1/3} \hskip 2 pt \mathrm{MeV}/c$ is the neutrons' Fermi momentum;
$R^D$, $R^M_n$ and  $R^M_p\leq1$ are reduction factors 
due to superfluidity~\cite{1994AstL...20...43L,yakovleven}, $a$~\cite{2009ApJ...707.1131P}, 
and $F$~\cite{1999A&A...343..650Y} are the dimensionless factor and the control function 
respectively both of which depend on the type of superfluidity. The factors in the above expressions which depend on 
superfluidity have been discussed briefly in Appendix~\re{superfluid}.
 
The Macro photon luminosity is 
\begin{equation}
    \mathcal L_{\gamma}=
    	10^{45}
    	\left(\frac{M_X/\rho_X}{M_\odot/\rho_N}\right)^{2/3}
    	\left({T^{\mathrm{s}}_{9}}^4 - {T^{\mathrm{CMB}}_{9}}^4\right) 
    	\mathrm{erg/s }
\label{photonL}
\end{equation}


where $T^{\mathrm{s}}$ is the surface temperature of the Macro,
and $T^{\mathrm{CMB}}$ is the temperature of the ambient plasma.

We assume that the Macros have coalesced, 
and we can begin following their cooling from   
when the temperature of the ambient plasma is $10^9$K,
at $z=3.7\times10^{8}$.
(This is after any electroweak and QCD-associated 
phase transitions~\cite{1990PhLB..240..179N}.)
We take the Macro to be isothermal at that moment
with temperature equal to that of the plasma.
The interior electrons, neutrons and protons will be degenerate.
The cooling of neutron stars below this temperature
has been well explored, 
and we have verified that
our conclusions are insensitive to the
details of the Macro cooling before this epoch.

We assume that the Macro, like a NS,
has a degenerate isothermal interior 
containing neutrons, protons and electrons,
and a non-degenerate ``atmosphere'' of electrons and heavy ions.
This keeps the interior warm as the ambient plasma cools.
For constant atmospheric photon luminosity $\mathcal{L}_\gamma$
the atmospheric density and temperature are related by

\begin{equation}
\label{density_atm}
\rho_\mathrm{atm} = 
	1.2\times10^{10} \rho_N 
	\left(\frac{\mu (M_{\mathrm{X}}/M_\odot)\mathrm{erg/s}}
		{Z(1+W)\mathcal{L}_\gamma}
	\right)^{1/2} T_9^{3.25}
\end{equation}
where $\mu$, $Z$ and $W$ are the mean molecular weight, metallicity or mass fraction of elements heavier than hydrogen and helium, and mass fraction of hydrogen. In case of a low metallicity atmosphere, the Kramer's opacity due to bound-free transitions assumed above will be exceeded by the opacity due to free-free transitions, which does not vanish for low metallicity. This means Eq.~\re{density_atm} will not diverge for small $Z$.

Where the atmosphere meets the interior,
the relation between density $\rho_{*}$ and temperature $T^*$ 
can be found (\cite{book} Chapter 4)
by equating the electron pressure 
of the degenerate interior and the non-degenerate atmosphere
\begin{equation}
\label{density*}
\rho_{*} = 7.6\times 10^{5}  
 \mu_{e}  {T^{*}_{9}}^{3/2}\,\,  \mathrm{g}/\mathrm{cm}^{3} \,.
\end{equation}
where $\mu_{e}$ is the mean molecular weight per electron. 
Equating~\re{density_atm} and~\re{density*},
\begin{equation}
\label{lumin}
\mathcal{L}_\gamma=
8.9\times 10^{36} ~
	\lambda ~
	(M_{\mathrm{X}}/M_\odot) ~
	{T^{*}_{9}}^{3.5}\,\,\mathrm{erg/s}\,,
\end{equation}
where 
	$\lambda\equiv\left(\frac{\mu}{\mu_{e}^{2}}\right)
				\frac{2.0}{Z(1+W)}$.
In the case of a NS, 
$\lambda\approx1$ (see~\cite{book}, Chapter 11).
We take $\lambda$ to be a free parameter that along with $\rho_{\mathrm{X}}$, 
represent the unknown characteristics of the Macros. 
We provide a brief derivation of the 
photon luminosity in Appendix~\re{Tmacro}.

The Macro interior is nearly isothermal, 
due to the thermal conductivity 
of the degenerate electrons. 
Since $T^{\mathrm{X}} \simeq T^{*}$,
equating the photon luminosity ~\re{photonL} at the surface
to~\re{lumin} yields the Macro's surface temperature
$T^{\mathrm{s}}_{9}(t)$. 

Starting from its assumed initial isothermal condition at $10^9$K,
the Macro cools according to 
\begin{equation}
\label{duxdt}
\frac{dU_{\mathrm{X}}}{dt} = 
	 - (\mathcal{L}^{\mathrm{DU}}_{\nu}
	 +\mathcal{L}^{\mathrm{MU}}_{\nu}
	 +\mathcal{L}^{\mathrm{CP}}_{\nu}
	 +\mathcal{L}_{\gamma}).
\end{equation}
where the internal energy is (\cite{book} Eq.(11.8.2))
\begin{equation}
 \label{intE}
U_{\mathrm{X}} = 
	6.1\times10^{47}
	\left(M_{\mathrm{X}}/M_\odot\right)  
	\left(\rho_{\mathrm{X}}/\rho_N\right)^{-2/3} 
	{T^{X}_{9}}^{2} \hskip 4 pt \mathrm{erg}.
\end{equation}
We refer the reader to Appendix~\re{Tmacro} for a derivation of the above equation. 

The interior temperature of the Macro therefore obeys
\be
\label{tmurcadurcacpphoton}
\frac{d}{dt}T^{\mathrm{X}}_{9}
	= -8.3\times10^{-4} s^{-1} 
		\left(\rho_X/\rho_N\right)^{1/3}
		{T^{X}_{9}}^{-1}
		\sum_i C_i
\ee
where the sum over $i$ now includes photons, and \newline
$C_\gamma\equiv 8.9\times10^{-9}
        \left(\rho_X/\rho_N\right)^{1/3}\lambda
        \left(T^{\mathrm{X}}_{9}\right)^{7/2}$.

Neutrino emission via MUrca occurs from the onset,
since we take the initial temperature to be $10^9$K.
Emission via CP begins below $T_{c9}$.
We explore two possibilities:
first, no DUrca cooling $R^D=0$;
second, a
proton fraction sufficient to support DUrca,
with $R^{D}$ given by Eq.19 in~\cite{1994AstL...20...43L}. 

In practice, SDs are relatively insensitive to the exact values of 
these various numerical factors.

In the case where there is negligible DUrca emission,
cooling proceeds in three stages:\newline
\noindent Stage 1:
		MUrca-dominated cooling from 
		$T^{\mathrm{X}}_{9}=T^{\mathrm{MU}}_{9}=1$,
		at time $t_9$,
		to $T^{\mathrm{CP}}_{9}=0.98T_{c9}$, at $t_{CP}$;\newline
\noindent Stage 2: 
		CP-dominated cooling from 
		$T^{\mathrm{CP}}_{9}$ 
		to $T^{\gamma}_{9}\simeq0.2T_{c9}$ at $t_{\gamma}$;\newline
\noindent Stage 3: 
		photon cooling below $T^{\gamma}_{9}$,
		i.e. after $t_{\gamma}$.

\begin{figure}
\includegraphics[width=0.65\hsize, angle=-90]{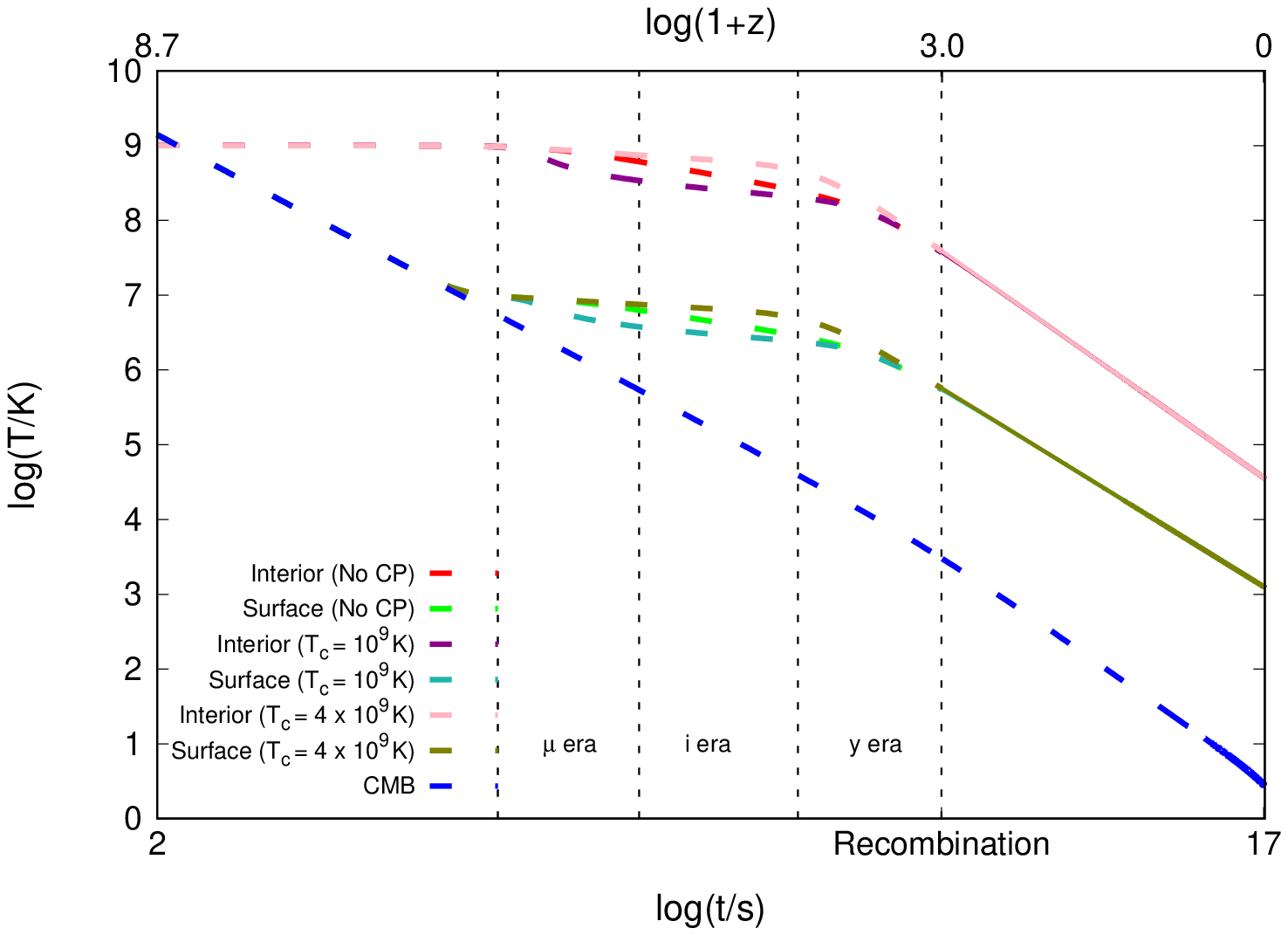}
\caption{\label{Tivst} Interior temperature $T_X$
and  surface temperature $T_s$ of a Macro for $M_X=M_\odot$, $\lambda=1$  and $\rho_X=\rho_N$,
plotted versus time $t$ and redshift $z$. 
Cooling is without DUrca, 
and both without CP and with CP for two values of $T_c$. 
The ambient photon temperature $T_{CMB}$ 
is shown for comparison, 
and the eras when $\mu$ and $y$ distortions occur are indicated.}
\end{figure}

(If $T^{CP}_{9}$ is high enough, the first stage may be omitted.)
The Macro cooling can be followed numerically, 
but by assuming that the dominant cooling mechanism in each stage
is the only one 
(and taking $R^{M}_{n}, R^{M}_{p}=1$), 
we find 
\begin{align}
	\label{Tx9oft_analytic}
	T^{\mathrm{X}}_{9}(t)\simeq& \\
	&\!\!\!\!\!\!\!\!\!\!\!\!\!\!\!\!\!\!
	\begin{cases}
		T^{\mathrm{MU}}_{9} 
			\left[1+2.7\times10^{-8}\alpha
			{T^{\mathrm{MU}}_{9}}^{6}
				\left(\frac{\rho_{\mathrm{X}}}{\rho_N}
					\right)^{1/3}
				\frac{t-t_{9}}{s}\right]^{-1/6} & \\
			\quad\quad\quad \mathrm{for}\,\, 
				1\equiv T^{\mathrm{MU}}_{9}
				\geq T^{\mathrm{X}}_{9}
				\geq T^{\mathrm{CP}}_{9}\simeq 0.98 T_{c9}\,, & \\
		T^{\mathrm{CP}}_{9}
			\left[1+1.8\times10^{-2} a  F 
			{T^{\mathrm{CP}}_{9}}^{5}
			~\frac{t-t_{\mathrm{CP}}}{s}\right]^{-1/5} &\\
			\quad\quad\quad \mathrm{for}\,\,
				0.98 T_{c9}\simeq T^{\mathrm{CP}}_{9}
				\geq T^{\mathrm{X}}_{9}
				\geq T^{\gamma}_{9}\simeq0.2 T_{c9}\,,\\
		T^{\gamma}_{9}
			\left[1.0+1.1\times10^{-11}\lambda
			{T^{\gamma}_{9}}^{3/2} 
			\left(\rho_{\mathrm{X}}/\rho_N
				\right)^{2/3}
			\frac{t-t_{\gamma}}{s}\right]^{-2/3}&\\
			\quad\quad\quad \mathrm{for}\,\,
			0.2 T_{c9} \simeq T^{\gamma}_{9}	\geq T^{\mathrm{X}}_{9}\,.
	\end{cases}\nonumber
\end{align}
The above relations can also be expressed in terms of redshift, $z$, using the time-redshift relation $z=4.9\times10^{9}(t/\mathrm{s})^{-1/2}$.
The times (and thus redshifts) 
at which the interior temperature $T_X$
falls to $T_{\gamma}$ 
depend on detailed properties of the Macro, 
such as its central density $\rho_X$, and the composition parameter $\lambda$. 
In Figure \ref{Tivst} we plot the central and surface
temperatures of the Macro,  as well as the CMB temperature as a function of time for a representative
value of these parameters.

In the presence of DUrca cooling,
Stage 1 is DUrca dominated until $T^{X}_{9}$ becomes $T^{\gamma}_{9}=0.1\hskip 2 pt T_{c9}$ at $t_{\gamma}$.
In this case, during Stage 1 (now $T^{\mathrm{DU}}_{9}\equiv1$)
\begin{align}
\label{DU}
T^{\mathrm{X}}_{9}(t)&=T^{\mathrm{DU}}_{9}\!\!
\left[1+0.017\left(\!T^{\mathrm{DU}}_{9}\!\right)^{\!4}R^{D}
\left(\!\frac{\rho_{\mathrm{X}}}{\rho_\mathrm{N}}\!\right)^{\!1/3}
\frac{t-t_{9}}{s}\right]^{-1/4}\!\!\!\!.
\end{align}
For convenience, we provide Table where we describe the various temperatures and times that appear in equations~\re{Tx9oft_analytic} and~\re{DU}.

\begin{table}[!h] 
\setlength{\tabcolsep}{2pt}
\renewcommand{\arraystretch}{1.5}
\begin{center}
\begin{tabular}{|c|c|}
\hline

& Description
\\
\hline
$T^{\mathrm{X}}_{9}$ & Interior temperature of macro
\\
\hline
$T^{\mathrm{CMB}}_{9}$ & CMB temperature
\\
\hline
$T^{\mathrm{MU}}_{9}$ & Temperature of macro at onset of MURCA
\\
\hline
$T^{\mathrm{DU}}_{9}$ & Temperature of macro at onset of DURCA
\\
\hline
$T^{\mathrm{CP}}_{9}$ & Temperature of macro at onset of CP cooling
\\
\hline
\end{tabular}
\end{center}
\vskip -0.3 cm
\caption{Definitions of various temperatures, $T^{i}_{9}$, that appear in equations~\re{Tx9oft_analytic} and~\re{DU}. For $i = $ MU, DU, CP, the temperature values were determined by comparing the luminosities of various processes given by Eq.~\re{luminosity}.}
\label{temp-table}
\end{table}
\begin{table}[!h] 
\setlength{\tabcolsep}{2pt}
\renewcommand{\arraystretch}{1.3}
\begin{center}
\begin{tabular}{|c|c|}
\hline

& Description
\\
\hline
$t_{9}$ & Cosmic time at which $T^{X}_{9}=1$
\\
\hline
$t_{\mathrm{CP}}$ & Cosmic time at which CP cooling dominates
\\
\hline
$t_{\gamma}$ & Cosmic time at which photon cooling dominates
\\
\hline
\end{tabular}
\end{center}
\vskip -0.3 cm
\caption{Definitions of various times, $t_{i}$, that appear in equations~\re{Tx9oft_analytic} and~\re{DU}. These time values very obtained by solving Eq.~\re{Tx9oft_analytic} for the most dominant cooling process.}
\label{time-table}
\end{table}

\section{Spectral Distortions by Macros}
\label{macros_sd}
The pre-recombination contributions 
to $\mu$ and $y$ distortions of the CMB can be approximated
by
\begin{equation}
\label{muy}
\left\{\begin{array}{c}\mu\\ y \end{array}\right\}
 = \int \hskip 2pt dt \hskip 2 pt \mathcal{J}_{\mathrm{bb}}  
 \left\{\begin{array}{c}
 	1.4 \mathcal{J}_{\mu} \\
 	\frac{1}{4}\mathcal{J}_{y}
 	\end{array}\right\}
	\frac{1}{c^{2}\rho_{\gamma}}\dot{Q}\,.
\end{equation}
The window functions given in~\cite{1970Ap&SS...7....3S},~\cite{1982A&A...107...39D},~\cite{2012JCAP...06..038K} are
\begin{eqnarray}
&&\mathcal{J}_{y}(z)=1-\mathcal{J}_{\mu}(z)\approx \left[1+4.7\times10^{-13}{z}^{2.58}\right]^{-1},\,\nonumber\\
&&\mathcal{J}_{\rm bb}(z)\approx  \exp\left[-(z/z_{\mu})^{5/2}\right] \,.
\end{eqnarray}
The CMB energy density
${\rho_{\gamma}}\approx 7.0\times10^{-34} z(t)^{4}~
	\mathrm{g/cm^{3}}$, 
while the rate at which energy density 
is injected into the photon distribution 
by Macros of density $n_X$ is
\begin{equation}
\dot{Q} = n_{\mathrm{X}} \mathcal{L}_{\gamma}.
\end{equation}
It is useful to rewrite 
$n_{\mathrm{X}} =\Omega_{\mathrm{X}0}\rho_{c}  z(t)^{3}/M_{\mathrm{X}}$
with 
$\rho_{c}\simeq10^{-29}\mathrm{g/cm^{3}}$,  
and Macro DM fraction $\Omega_{\mathrm{X}0}\lsim0.24$.

\begin{figure*}
\centering
  \includegraphics[width=0.23\textwidth,angle=-90]{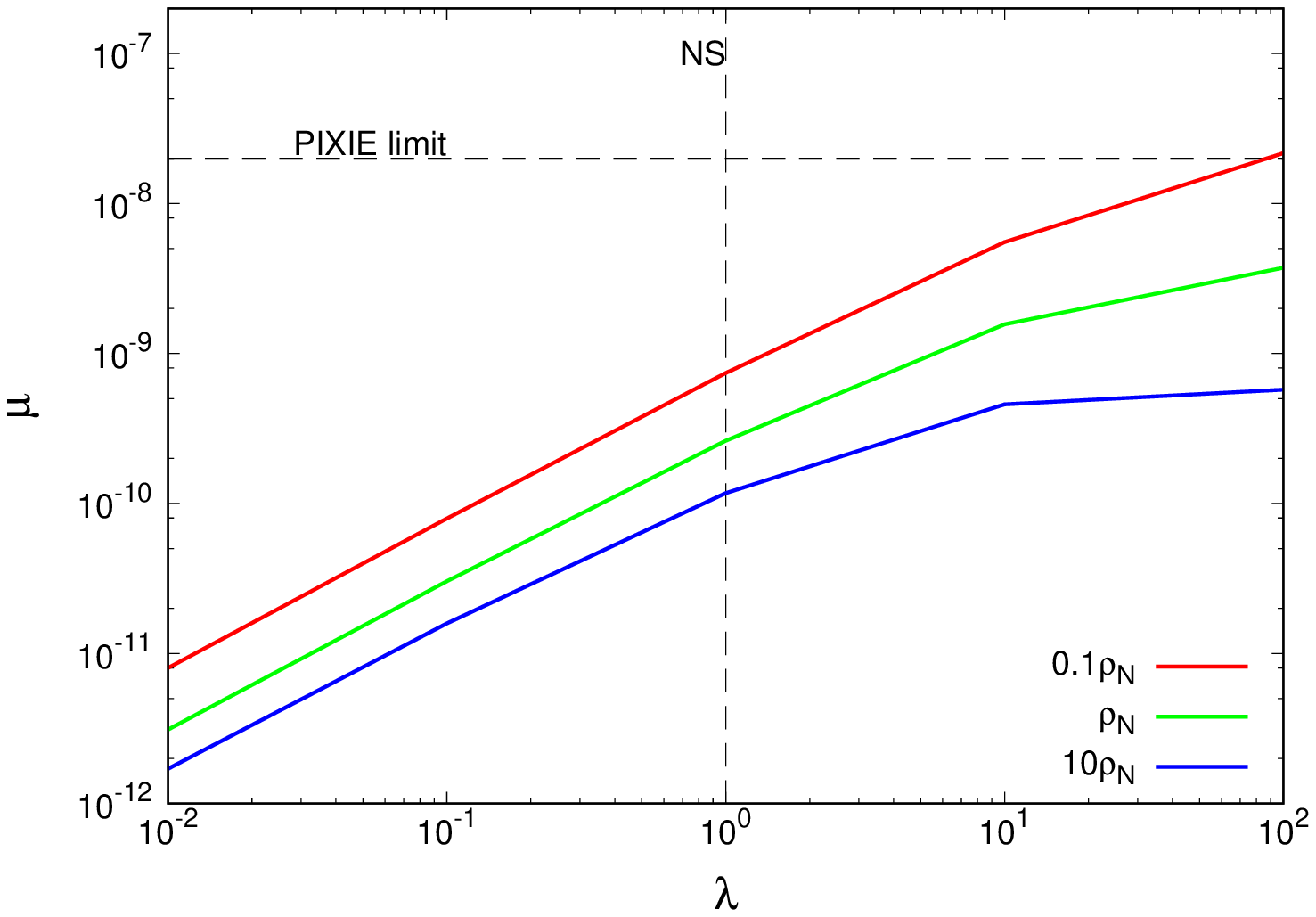}
  \includegraphics[width=0.23\textwidth,angle=-90]{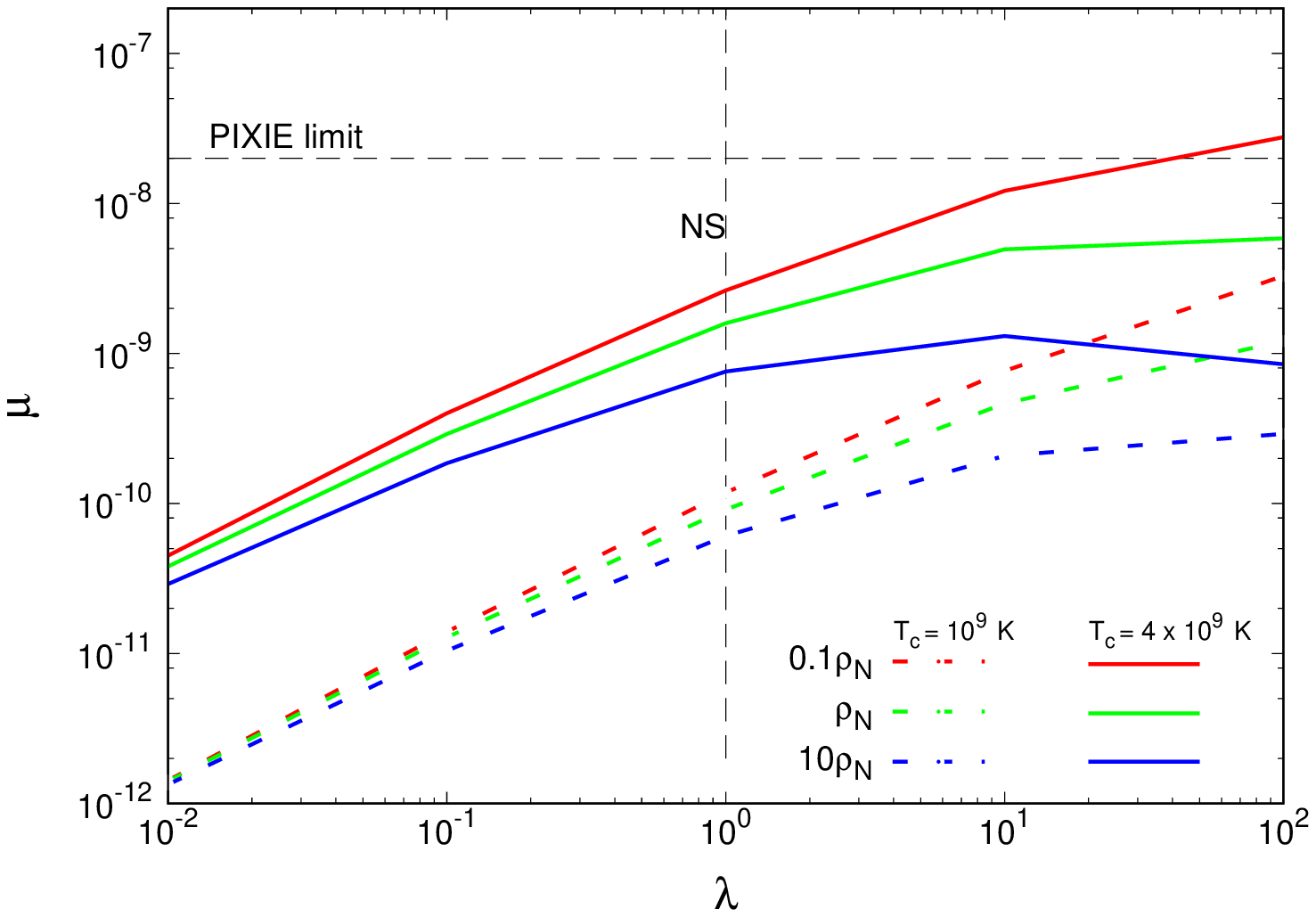}
  \includegraphics[width=0.23\textwidth,angle=-90]{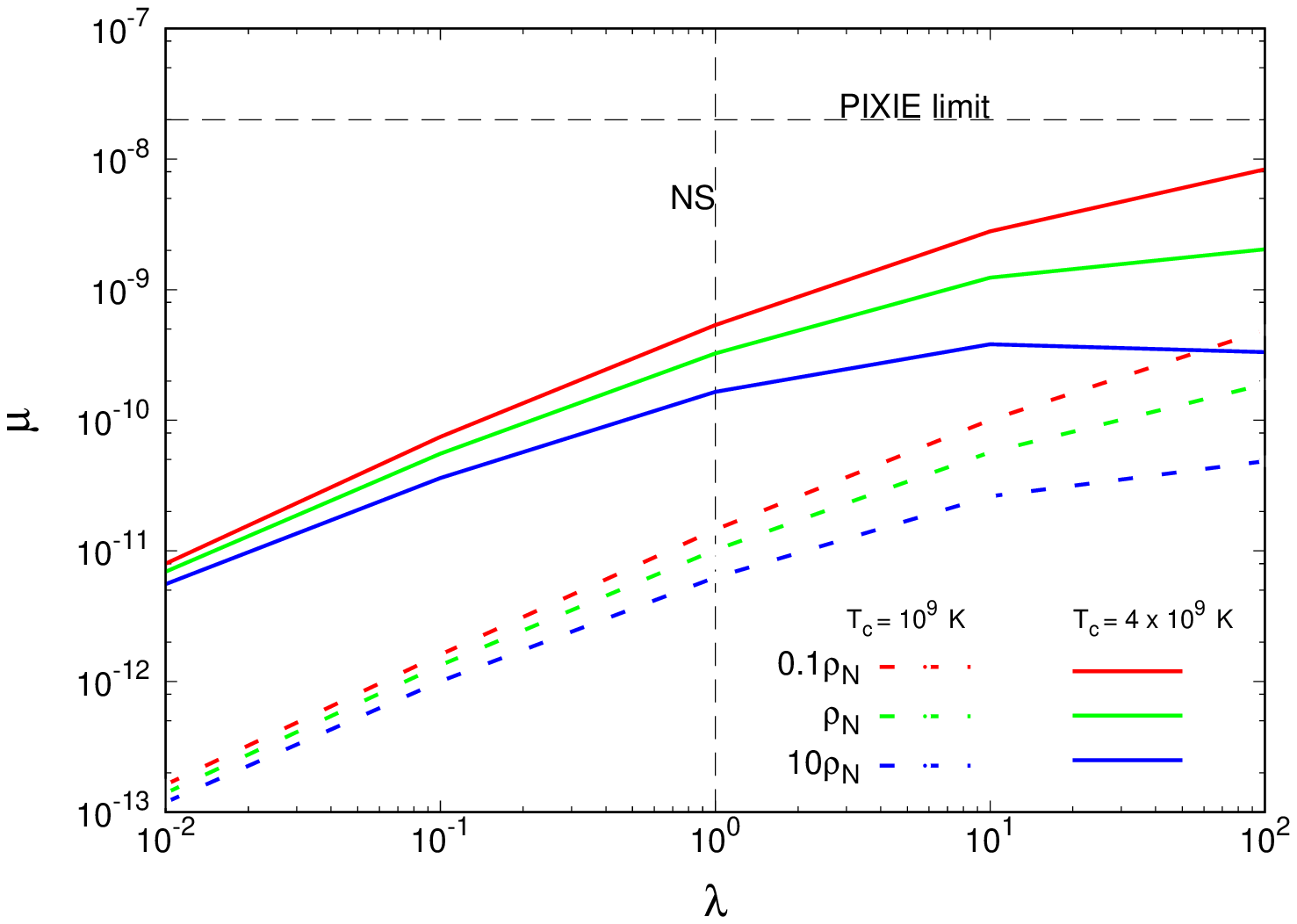}
  \centering
\includegraphics[width=0.23\textwidth,angle=-90]{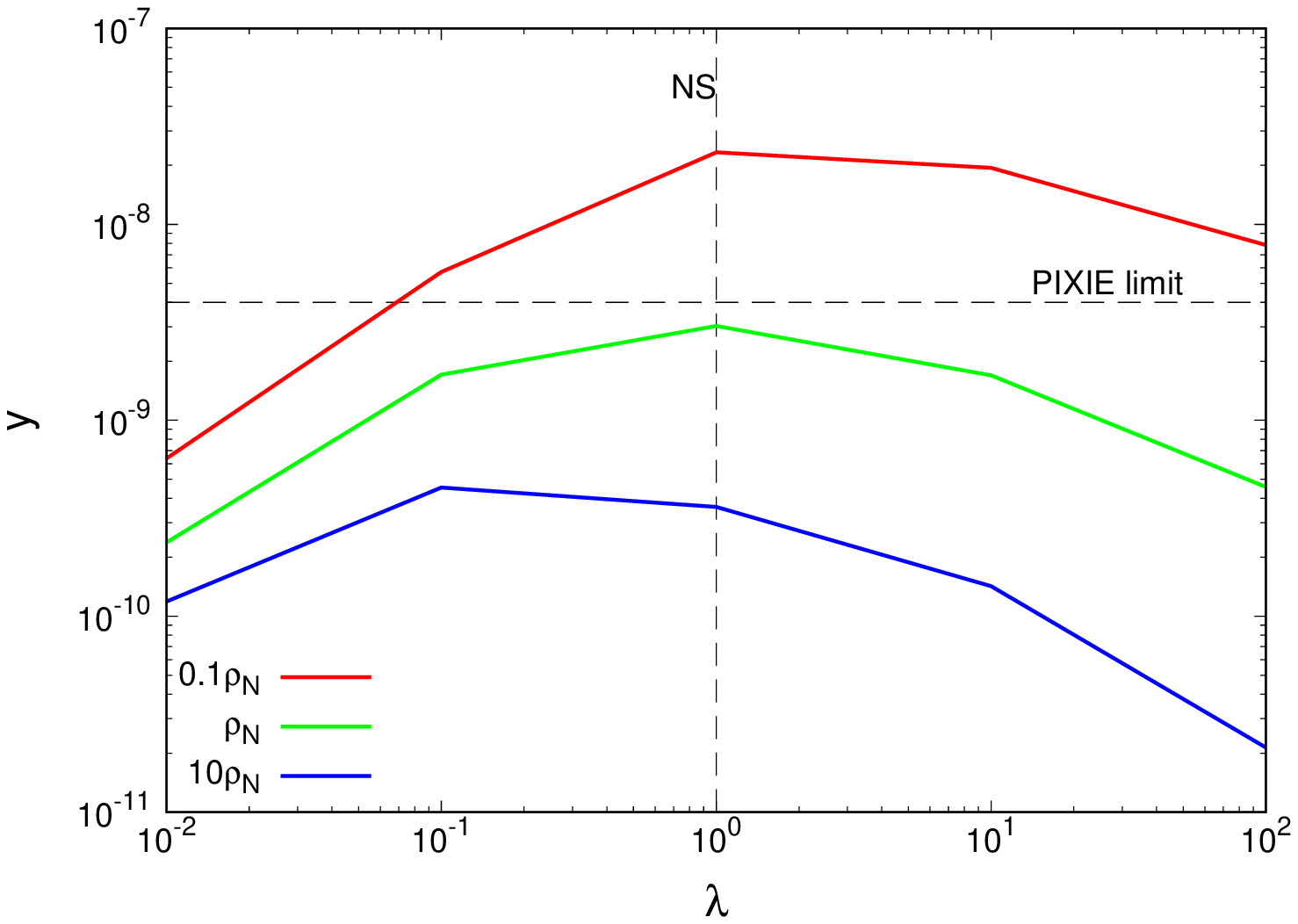}
\includegraphics[width=0.23\textwidth,angle=-90]{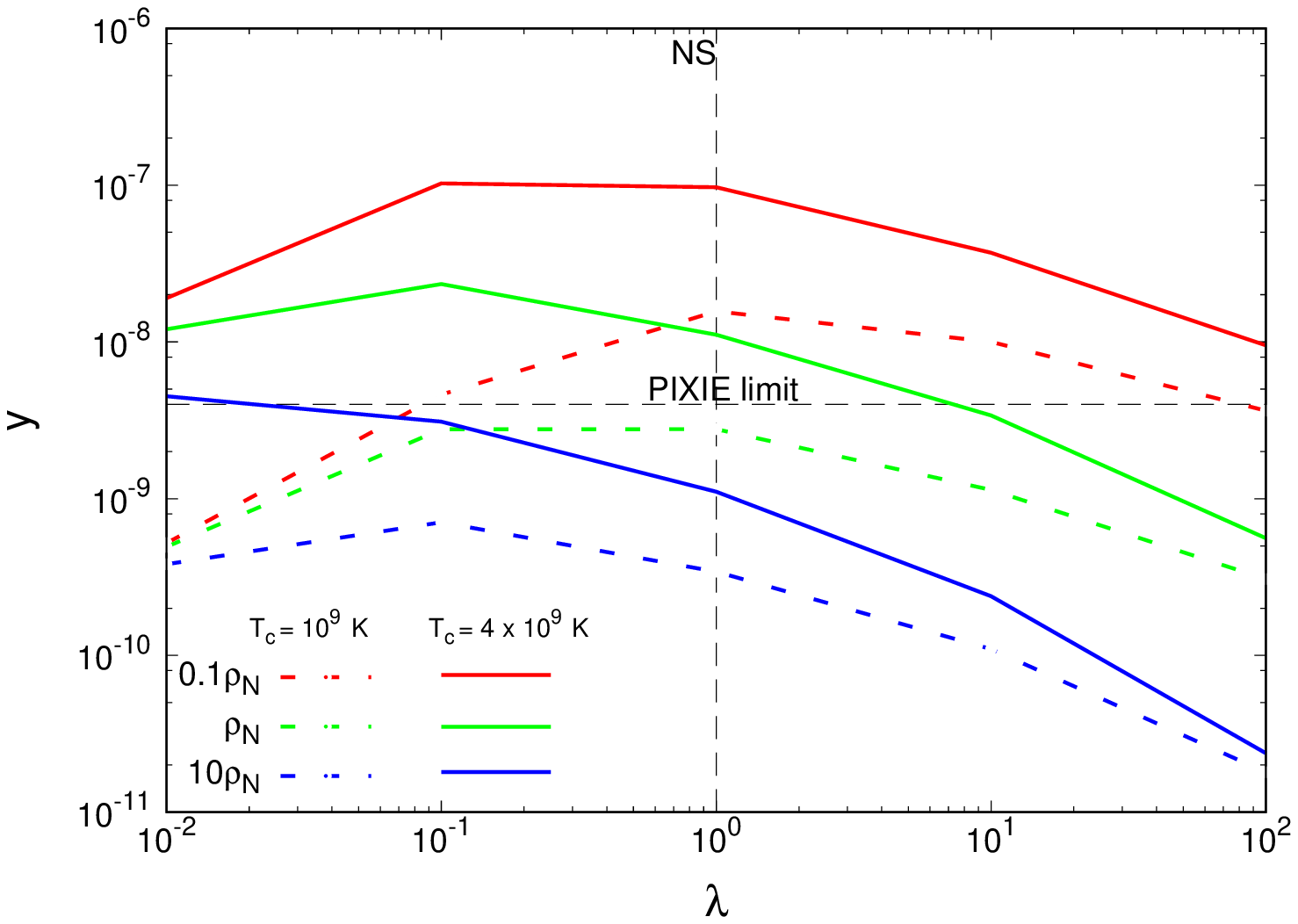}
\includegraphics[width=0.23\textwidth,angle=-90]{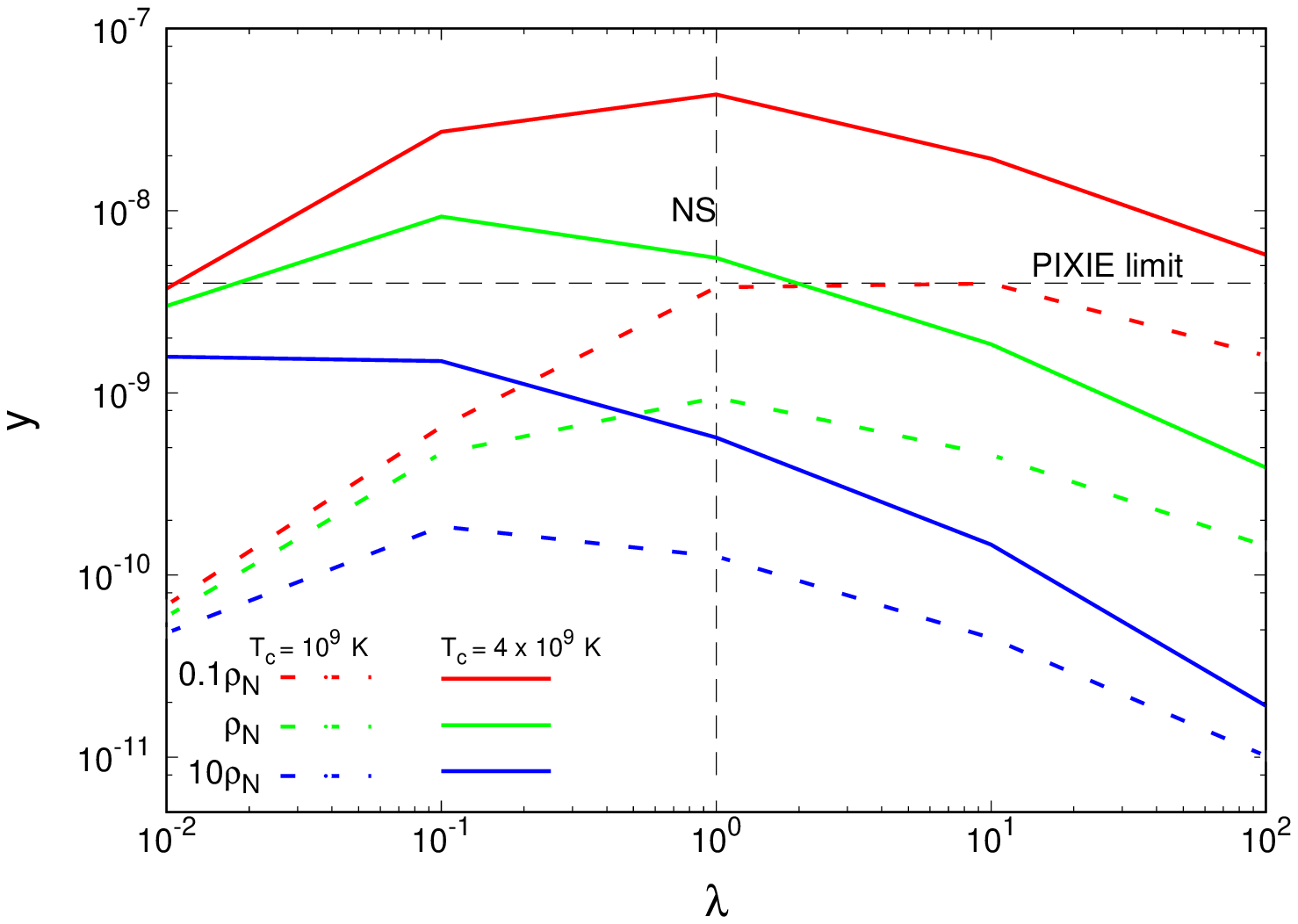}
  \caption{\label{muofl} Top panel: $\mu$ distortion as a function of Macro surface composition factor $\lambda$ for three different cooling scenarios.  On the left, no DUrca, no CP; in the 
  middle no DUrca, with CP; on the right; with DUrca, with CP.
  Green lines denote $\rho_X=\rho_N$, red lines $0.1\rho_N$; blue
  lines $10\rho_N$.
  The panels with CP show results for $T_c=10^9$K (dashed) 
  and $T_c=4\times10^9$K (solid). Bottom panel: as for top panel but for $y$ distortion. The vertical dashed line stands for $\lambda=1$ which is similar to a Neutron Star.}
\end{figure*}

In Figure \ref{muofl}, 
we plot $\mu$ and $y$ (obtained numerically) {\it vs.} $\lambda$ 
for 
$M_X=M_\odot$,
Macro densities near the fiducial $\rho_N$,
and a variety of cooling scenarios.

We have also calculated the perturbation to the neutrino
energy density, since neutrinos are injected well after weak-interaction freezeout at $T_{CMB9}\simeq 10$,
but the change in $N_{eff}$ is negligible.
This could change if the internal physics of the Macro were
radically different.

The predicted  $y$ distortion is comparable 
to the target sensitivities
of anticipated next-generation 
spectral distortion satellite missions,
and the predicted $\mu$ distortion is nearly so. 
We remind the reader that since  
Macros are much hotter than the plasma 
through much of their history
and stay hot well after recombination, 
$\mu$ and $y$ do not adequately capture
the detectability of the SD signal.

Although we have presented results for $M_X=M_\odot$,
the spectral distortions $\mu$ and $y$ are mass independent,
for fixed $\rho_X$, since ${\cal L}_\gamma\propto M_X$
from \eqref{lumin},
while $n_X\propto M_X^{-1}$, and so $\dot{Q}\propto M_X^0$.
The distortions will however depend on $\rho_X$,
as well as on the detailed physics of the surface layer
(as parametrized by $\lambda$), 
and the cooling mechanisms operative in an actual Macro.

\section{Conclusions}
\label{conclusions}
We have demonstrated that the presence of macroscopic DM in the early universe may lead to observable signatures 
in the CMB spectrum. To fully characterize these distortions, the full spectral distortion must be inferred numerically
using the Boltzmann equation  -- this includes so-called intermediate distortions, 
a more complete characterization of the distorted spectrum
and continued contributions to the distortions post-recombination. 
Also, the temperature of Macros post-recombination 
may stay much higher compared to CMB
for an extended period, 
implying the presence of hot relics that could be visible as an  associated background radiation, 
or could heat the  post-recombination universe.

Other signatures can also be anticipated,
such as correlations between CMB temperature anisotropies and
spectral distortion anisotropies,
the presence of heavy elements in the pre-recombination universe,
and the continued production of these elements post-recombination
and outside stars. 
The unexplored possibilities for observable consequences of standard model
DM are yet rich.

\acknowledgments
SK, GDS and BWL are partially supported by US DoE grant DE-SC0009946. 

\appendix
\section{Photon Luminosity and Internal Temperature of Macro}
\label{Tmacro}

In this Appendix, we derive the temperature dependence of the Macro given by Eq.~\re{tmurcadurcacpphoton}, following closely the treatment of Chapter 4 in~\cite{book}. 

We assume that below degeneracy temperature of $10^{9}$  K, the core of the Macro is composed of degenerate neutron-proton-electron plasma, and is isothermal. The atmosphere is composed of a non-degenerate layer. The energy transfer due to photon diffusion from the hot interior to the ambient CMB through the atmosphere can be described by the radiative heat transfer equation assuming local thermal equilibrium and steady state. The photon luminosity, $\mathcal{L}_{\gamma}$, is given by
\be
\label{rht}
\mathcal{L}_{\gamma} = -4\pi r^{2} \frac{c}{3\kappa\rho_\mathrm{atm}}\frac{d}{dr}(a T^{4}),
\ee
where $a$ is the radiation constant, $r$ is the radial distance from the center of the Macro, $\kappa$, $\rho_\mathrm{atm}$ and $T$ are the Rosseland mean opacity, density, and the temperature of the atmosphere respectively. 

Opacity, $\kappa$, can be approximated as Kramer's opacity 
\be
\label{kramer}
\kappa = \kappa_{0} \rho_\mathrm{atm} T^{-3.5},
\ee
where
\be
\kappa_{0} = 4.34\times 10^{24} Z(1 + X) \hskip 2pt  \mathrm{cm}^{2}  \mathrm{g}^{-1}.
\ee
Hydrostatic equilibrium  requires that the pressure of the atmosphere depends on the radius as
\be
\label{hydrostatic}
\frac{dP}{dr} = -\frac{G m(r) \rho_\mathrm{atm}}{r^{2}},
\ee
where $m(r)$ is the mass of the Macro within the radius $r$. Since the atmosphere is much thinner than the radius of the core, we can set $m(r) = M_{X}$. 

The pressure for a non-degenerate gas is also given by the ideal gas law:
\be
\label{ideal}
P(r)=\frac{\rho_\mathrm{atm}}{\mu m_{u}}k_{B} T,
\ee
where $\mu m_{u}$ is the mean molecular weight. ($m_{u}$ is the atomic mass unit.)

Substituting~\re{ideal} in~\re{hydrostatic}, and using~\re{kramer} and~\re{rht},
\be
P \hskip 2 pt dP = 5.33ac\frac{\pi G M_{X}}{\kappa_{0}\mathcal{L}_{\gamma}}\frac{k_{B}}{\mu m_{u}}T^{7.5} dT. 
\ee 
Assuming a constant luminosity throughout the thin atmosphere, we can integrate the above equation with the boundary condition, $P=0$ when $T=0$. Thus we arrive at the density of the atmosphere given by Eq.~\re{density_atm}:
\begin{eqnarray}
\label{rhoatm}
\rho_\mathrm{atm} &&= \sqrt{1.25 a c \frac{\pi G M}{\kappa_{0}\mathcal{L}_{\gamma}}\frac{\mu m_{u}}{k_{B}}}T^{3.25} \\
&&=	1.2\times10^{10} \rho_N 
	\left(\frac{\mu (M_{\mathrm{X}}/M_\odot)\mathrm{erg/s}}
		{Z(1+W)\mathcal{L}_\gamma}
	\right)^{1/2}T_9^{3.25}.\nonumber 
\end{eqnarray}

At the point where the atmosphere meets the core, the non-degenerate electron pressure of the atmosphere given by the ideal gas law is equal to the  electron degeneracy pressure of the core,  
\be
\frac{\rho_{*}k_{B}T_{*}}{\mu_{e}m_{u}}=1.0\times10^{13}\left(\frac{\rho_{*}}{\mu_{e}}\right)^{5/3},
\ee
where $\mu_{e}$ is the mean molecular weight per electron, and $\rho_{*}$ and $T_{*}$ are the density and temperature at this transition point. Solving for $\rho_{*}$ in the above equation and equating it with~\re{rhoatm}, we get the luminosity of the Macro given by Eq.~\re{lumin}:
\begin{eqnarray}
\label{photonapp}
\mathcal{L}_{\gamma}&&=\left(5.7\times10^{5} \mathrm{erg/s}\right)\frac{\mu}{\mu_{e}^{2}}\frac{1}{Z(1+W)}\frac{M_{\mathrm{X}}}{M_{\odot}}T_{*}^{3.5}\nonumber \\
&&=(8.9\times 10^{36}\mathrm{erg/s}) ~
	\lambda ~
	(M_{\mathrm{X}}/M_\odot) ~
	{T^{*}_{9}}^{3.5}.\,\,
\end{eqnarray}

From~\cite{1939isss.book.....C}, the heat capacity of the Macro at temperature $T_{\mathrm{X}}$ is 
\be
C_{v} = \frac{dU_{\mathrm{X}}}{dT_{\mathrm{X}}}\bigg{|}_{N,V}=\frac{\pi^{2}(x^{2} + 1)^{1/2}}{x^{2}}N k_{B} \left(\frac{k_{B}T_{\mathrm{X}}}{m c^{2}}\right),
\ee
where $U_{\mathrm{X}}$, $N$, and $V$ are the internal energy, total number of neutrons, and volume of the Macro respectively. In the above equation, $x=p_{f}/m_{n}c$is the relativity parameter, where $p_{f}$ is the Fermi momentum, and $m_{n}$ is the mass of neutron. Integrating the above equation over $T_{\mathrm{X}}$ gives us Eq.~\re{intE} for the internal energy:
\begin{equation}
U_{\mathrm{X}} = 
	(6.1\times10^{47}~\mathrm{erg})
	\left(\frac{M_{\mathrm{X}}}{M_\odot}\right)  
	\left(\frac{\rho_{\mathrm{X}}}{\rho_N}\right)^{-2/3} 
	{\left(T^{X}_{9}\right)}^{2}.
\end{equation}

We can use this expression for $U_{\mathrm{X}}$ in the left hand side of Eq.~\re{duxdt}. The right hand side of Eq.~\re{duxdt} is a sum of the photon luminosity \ref{photonapp}, and neutrino luminosities that we will briefly describe in the appendix below.

\section{Neutrino Emission Luminosity}
\label{superfluid}

In this appendix, we describe briefly the neutrino emissions from the Macro as given by Eq.~\re{luminosity}. A detailed derivation of~\re{luminosity} is beyond the scope of the paper. Moreover, the expressions for the luminosities are very well established, and have been studied in great detail \cite{1994AstL...20...43L,yakovleven,1999A&A...343..650Y}.

The DURCA luminosity \cite{1994AstL...20...43L}
\begin{equation}
\mathcal{L}^{\mathrm{DU}}_{\nu} = 
	5.2\times10^{45}~(T_{9}^{X})^6~R^{D}\left(\frac{M_{\mathrm{X}}}{M_\odot}\right)
	\left(\frac{\rho_{\mathrm{X}}}{\rho_N}\right)^{-1/3}
	 \mathrm{erg/s}
\end{equation}
where $R^{D}$ is the reduction factor in DURCA rate due to superfluidity. As an example, we considered the type-AA superfluidity of neutrons and protons. The $R^{D}$ is given by Eq.19 in~\cite{1994AstL...20...43L}:
\begin{equation}
R^{D} = \frac{u}{u+0.9163}S + D,\,\nonumber\\
\end{equation}
\begin{equation}
S = \frac{1}{I_{0}}(K_{0}+K_{1}+0.42232K_{2})\left(\frac{\pi}{2}\right)^{1/2}p_{s}^{1/4}e^{-\sqrt{p_{e}}},\,\nonumber\\
\end{equation}
\begin{equation}
I_{0} = 457\pi^{6}/5040,\nonumber\\
\end{equation}
\begin{eqnarray}
&&K_{0}=\frac{\sqrt{p-q}}{120}(6p^{2} + 83pq + 16q^{2})\nonumber \\
&& - \sqrt{p}\frac{q}{8}(4p+3q) \mathrm{ln}\left(\frac{\sqrt{p}+\sqrt{p-q}}{\sqrt{q}}\right),\,\nonumber
\end{eqnarray}
\begin{equation}
K_{1} = \frac{\pi^{2}\sqrt{p-q}}{6}(p+2q)-\frac{\pi^{2}}{2}q\sqrt{p}\mathrm{ln}\left(\frac{\sqrt{p}+\sqrt{p-q}}{\sqrt{q}}\right),\,\nonumber\\
\end{equation}
\begin{equation}
K_{2} = \frac{7\pi^{4}}{60}\sqrt{p-q},\nonumber \\
\end{equation}
\begin{equation}
2p = u + 12.421 + \sqrt{w^{2} + 12.350u + 45.171},\nonumber \\
\end{equation}
\begin{equation}
2q = u + 12.421 - \sqrt{w^{2} + 12.350u + 45.171},\nonumber \\
\end{equation}
\begin{equation}
2p_{s} = u + \sqrt{w^{2} + 5524.8u + 6.7737},\nonumber \\
\end{equation}
\begin{equation}
2p_{e} = u + 0.43847 + \sqrt{w^{2} + 8.3680u + 491.32},\nonumber \\
\end{equation}
\begin{equation}
D = 1.52(u_{1}u_{2})^{3/2}(u_{1}^{2}+u_{2}^{2})e^{-u_{1}-u_{2}},\nonumber \\
\end{equation}
\begin{equation}
u_{1}=1.8091+\sqrt{v_{1}^{2}+(2.2476)^{2}},\nonumber \\
\end{equation}
\begin{equation}
u_{2}=1.8091+\sqrt{v_{2}^{2}+(2.2476)^{2}},\nonumber \\
\end{equation}
\begin{equation}
u = v_{1}^{2} + v_{2}^{2},\nonumber \\
\end{equation}
\begin{equation}
w = v_{2}^{2} - v_{1}^{2},\nonumber \\
\end{equation}
\begin{equation}
v_{1} = v_{2} = v_{A} = \sqrt{1-\tau}\left(1.456-\frac{0.157}{\sqrt{\tau}}+\frac{1.764}{\tau}\right), 
\end{equation}
where 
\be
\tau \equiv \frac{T_{X}}{T_{c}}.
\ee 

The MURCA luminosity \cite{yakovleven}
\begin{eqnarray}
\mathcal{L}^{\mathrm{MU}}_{\nu} && = (3.0R^{M}_{{n}}+2.4R^{M}_{{p}})10^{39} ~(T_{9}^{X})^8~\alpha \nonumber\\
	&&\times\left(M_{\mathrm{X}}/M_\odot\right)
	\left(\rho_{\mathrm{X}}/\rho_N\right)^{-1/3}
	 \mathrm{erg/s} 
\end{eqnarray}
For simplicity, we consider only singlet-state neutron superfluidity of Type-\textbf{A}.
The associated reduction factors $R^{M}_{n}$ and $R^{M}_{p}$ are given by Eq.~32 and Eq.~37 in~\cite{yakovleven}
\begin{equation}
	R^{M}_{n} = \frac{a^{7.5} + b^{5.5}}{2}\hskip 2 pt e^{3.4370 - \sqrt{(3.4370)^{2} + v^2}},\nonumber \\
\end{equation}
\begin{equation}
	a = 0.1477 + \sqrt{(0.8523)^{2} + (0.1175v)^2},\nonumber \\
\end{equation}
\begin{equation}
	b = 0.1477 + \sqrt{(0.8523)^{2} + (0.1297v)^2}, \nonumber \\
\end{equation}
\begin{equation}
	v = \sqrt{1 - \tau}\left(1.456 - \frac{0.157}{\sqrt{\tau}} + \frac{1.764}{\tau}\right), \nonumber \\
\end{equation}
and
\begin{eqnarray}
R^{M}_{p} &&= \left(0.2414 + \sqrt{(0.7586)^{2} + (0.1318 v)^{2}}\right)^{7}\nonumber \\
&& e^{5.339 - \sqrt{(5.339)^{2} + (2 v)^{2}}}.
\end{eqnarray}
respectively. 

The CP cooling luminosity \cite{1999A&A...343..650Y}
\begin{eqnarray}
\mathcal{L}^{\mathrm{CP}}_{\nu} &&=   7.1\times10^{39} \mathrm{erg/s}~(T_{9}^{X})^7~a ~F\\ \nonumber
&&\times \left(M_{\mathrm{X}}/M_\odot\right) \left(\rho_{\mathrm{X}}/\rho_N\right)^{-2/3}  \mathrm{erg/s} 
\end{eqnarray}
The function $F$ controls the efficiency of the CP cooling process. We select  $F$ to be $F_{A}$ given by
\begin{eqnarray}
F_{A}(v) && = (0.602 v^{2} + 0.5942 v^{4} + 0.288 v^{6}) \nonumber \\
&& \times \left(0.5547 + \sqrt{(0.4453)^{2} + 0.01130 v^{2}}\right)^{1/2} \nonumber \\
&& \times e^{-\sqrt{4 v^2 + (2.245)^{2}} + 2.245}
\end{eqnarray}
 Eq.(34) in~\cite{1999A&A...343..650Y}. The factor $a$ in the CP luminosity is a constant that depends on nucleon species and superfluidity type. It has the maximum value of $4.17$ and $3.18$ for triplet states of neutrons and protons respectively.

\bibliography{references}
\end{document}